\documentclass{article}
\usepackage{fullpage}

\usepackage{graphicx}
\usepackage{subcaption}
\usepackage{float}
\usepackage{natbib}
\usepackage{amssymb}
\usepackage{amsmath}
\usepackage{trimclip}
\usepackage{color}
\usepackage[section]{placeins}
\usepackage{upgreek}
\usepackage{relsize,exscale}
\usepackage{bm}
\usepackage{xcolor}

\usepackage[normalem]{ulem}
\usepackage{soul}

\def\Dlt{\triangle}

\def\i{\mathrm{i}}
\def\bcdot{\bm{\cdot}}

\def\e{\mathrm{e}}
\def\eps{\varepsilon}
\def\d{\mathrm{d}}

\newcommand{\Ro}{\mbox{\textit{Ro}}}
\newcommand{\Bu}{\mbox{\textit{Bu}}}

\newcommand\barL[1]{\overline{#1}^\mathrm{L}}

\newcommand\barS[1]{\overline{#1}^\mathrm{S}}

\def\beq{\begin{equation}}
\def\eeq{\end{equation}}
\newcommand{\com}{\, ,}
\newcommand{\per}{\, .}

\newcommand{\bxi}{\boldsymbol{\xi}}
\newcommand{\ddxi}{(\bxi' \bcdot \grad)}

\newcommand{\cross}{\times}
\newcommand{\grad}{\boldsymbol{\nabla}}
\newcommand{\curl}{\boldsymbol{\nabla} \!\times\!}
\renewcommand{\div}{\boldsymbol{\nabla} \bcdot}

\newcommand{\bu}{\ensuremath {{\boldsymbol {u}}}}

\newcommand{\half}{\tfrac{1}{2}}
\newcommand{\bh}{\ensuremath {\boldsymbol {h}}}

\newcommand\WRY[1]{{\color{violet} #1}}

\title{Wave-averaged  balance:  a simple example} 

\author{Hossein A. Kafiabad$^1$, Jacques Vanneste$^1$ and William  R. Young$^2$}   
\date{\normalsize{$^1$School of Mathematics and Maxwell Institute for Mathematical Sciences, \\
University of Edinburgh, Edinburgh, UK} \\
$^2$Scripps Institution of Oceanography, University of California, San Diego, USA}

\begin{document}

\maketitle

\begin{abstract}
In the presence of inertia-gravity waves, the geostrophic and hydrostatic balance that characterises the slow dynamics of rapidly rotating, strongly stratified flows holds in a time-averaged sense and applies to the Lagrangian-mean velocity and buoyancy. We give an elementary derivation of this wave-averaged balance and illustrate its accuracy in numerical solutions of the three-dimensional Boussinesq equations, using a simple configuration in which vertically planar near-inertial waves interact with a barotropic anticylonic vortex. We further use the conservation of the wave-averaged potential vorticity to predict the change in the barotropic vortex induced by the waves. 
\end{abstract}

\section{Introduction}\label{sec:intro}

Our understanding of the large-scale dynamics of the atmosphere and ocean rests on the concept of geostrophic balance: motion with timescales much longer than the inertial timescale $f^{-1}$, with $f$ the Coriolis parameter, is assumed to dominate, resulting in the familiar balance between Coriolis force, buoyancy  and pressure gradients. The idea of dominant slow motion has stimulated decades of research extending balance beyond geostrophy to define dynamics which filter out fast motion to a high degree of accuracy -- that is, dynamics restricted to a suitably defined slow manifold \citep[e.g.][]{machenhauer1977dynamics,leith1980nonlinear,allen1993intermediate,warn1995rossby,mcintyre2000potential,mohebalhojeh2001hierarchies,vanneste2013balance,kafiabad2018spontaneous,kafiabad2017rotating}.  However, the amplitude of the fast motion, e.g.\ in the form of inertia-gravity waves, can often be large, corresponding to states well away from the slow manifold. For instance, the velocities associated with  near-inertial waves (NIWs hereafter) in the ocean frequently exceed those of the (vortical) balanced motion \citep[e.g.][]{alford2016near}. There is, therefore, a need to reassess the notion of balance to account for the impact of waves. This can be achieved using reduced models which, rather than filtering the waves, average over their rapidly oscillating phases to describe the slow  interactions between waves and vortical flow. We refer to these  as \textit{wave-averaged} models. 

\citet{bretherton1971general}, \citet{grimshaw1975nonlinear} and, more recently, \citet{wagner2015available} showed that wave averaging naturally leads to a generalised-Lagrangian-mean (GLM) description of the flow, in which averaging is carried out at fixed particle label rather than fixed Eulerian position.
Wave-averaged models can therefore be interpreted as approximations to the GLM equations of \citet{andrews1978exact} (see B\"uhler \citeyear{buhler2014waves} for an introduction to GLM). The wave-averaged models of interactions between NIWs  and quasigeostrophic flow derived by \citet{xie2015generalised} and \citet{wagner2016three} fall into this category. \cite{salmon2016variational} provides an interesting variational perspective into  this class of models in which weakly nonlinear internal waves interact perturbatively with quasigeostrophic flow.

One of the most striking predictions of wave-averaged and GLM theories is that a form of hydrostatic and geostrophic balance continues to hold in the presence of strong waves, but that this balance applies to the Lagrangian mean flow,  in the sense that
\begin{equation}
f \bm{z} \times \barL{\bm{u}} = - \grad \overline{\pi} + \barL{b} \bm{z},
\label{eq:averagedgeostrophy}
\end{equation}
where $\bm{z}$ is the vertical unit vector, $\barL{\bm{u}}$ and $\barL{b}$ are the Lagrangian-mean velocity and buoyancy, and $\overline{\pi}$ is a mean pressure-like scalar that includes a quadratic wave-averaged contribution \citep{moore1970mass,xie2015generalised,wagner2015available,gilbert2018geometric,thomas2018wave}.  The balance in \eqref{eq:averagedgeostrophy} follows   expeditiously  by simplification of the exact  GLM momentum  equation (\cite{andrews1978exact} Theorem I)  in the small Rossby number limit.  In section \ref{sec:balance} we provide an alternative derivation of \eqref{eq:averagedgeostrophy} proceeding  from the Eulerian equations.

 The Lagrangian-mean velocity, {$\barL{\bm{u}} $ in \eqref{eq:averagedgeostrophy},}  is the sum of Eulerian-mean and Stokes velocities,
\begin{equation} \label{eq:stokes}
 \barL{\bm{u}} = \overline{\bm{u}} + \barS{\bm{u}}, \quad \textrm{where} \ \ \barS{\bm{u}} = \overline{\bm{\xi}' \bcdot \grad \bm{u}'},
\end{equation}
with $\bm{\xi}'$ and $\bm{u}'=\bm{\xi}'_t$   denoting wave displacement and velocity and the overline denoting (Eulerian) time averaging.
The  appearance of the Stokes velocity $ \barS{\bm{u}}$  in the Coriolis term implies the existence of the Stokes--Coriolis force $f \bm{z} \times \barS{\bm{u}}$ whose importance for shallow currents driven by surface gravity waves has long been recognised \citep[e.g.][]{ursell1950theoretical,hasselmann1970wave,huang1979surface,leibovich1980wave,buhler2014waves}. Here we are concerned with the Stokes--Coriolis  force associated with  internal gravity waves which  operates in the interior ocean.
Equation \eqref{eq:averagedgeostrophy} is a remarkable generalisation of the familiar geostrophic balance, with practical implications, e.g.\ for the interpretation of ocean velocities inferred from satellite altimetry data, yet \eqref{eq:averagedgeostrophy} is not widely known. Another prediction of wave-averaged and GLM theories is the material conservation of a form of potential vorticity (PV) which combines the PV of the mean flow with a wave-averaged  contribution. Together with \eqref{eq:averagedgeostrophy} this leads to a wave-averaged form of quasigeostrophic dynamics which captures the feedback exerted by waves  on the mean flow.

The aim of this paper is to {illustrate} the validity and usefulness of the wave-averaged geostrophic balance and PV conservation by testing these relations against numerical solutions of the  three-dimensional Boussinesq equations. Applying wave-averaged relations to numerical simulations poses difficulties related to the definition of a suitable time average  and the estimation of particle displacements $\bm{\xi}'$ in \eqref{eq:stokes}. We sidestep these \WRY{issues} by focussing on a simple configuration: an NIW, initially with no horizontal dependence, and the vertical structure of a plane travelling wave, is superimposed on an axisymmetric (Gaussian) barotropic anticyclonic vortex. This configuration has the advantage that, because the NIW is approximately linear, it retains a simple vertical and temporal structure proportional to $\e^{\i(mz-ft)}$, with $m$ the vertical wavenumber. Time averaging can therefore be replaced by a straightforward vertical averaging and the Eulerian-mean component of the flow can be identified with the barotropic component. 
Moreover, it turns out that for NIWs, the Stokes drift $\barS{\bm{u}}$ can be deduced from the wave kinetic energy, thus circumventing the need to estimate the displacements $\bm{\xi}'$.

We describe our numerical simulation of the interaction between NIWs 
and an anticyclonic vortex in \S\ref{sec:model}. The wave evolution follows a broadly understood scenario: after a rapid adjustment, wave energy concentrates in the vortex core, giving rise to an approximately axisymmetric trapped structure which is modulated periodically in time \citep{llewellyn1999near}. Our focus is on the response of the mean flow to this wave pattern. In \S\ref{sec:balance}, we briefly sketch a derivation of the wave-averaged  balance \eqref{eq:averagedgeostrophy} and of the wave-averaged PV. Particularising this to the case of vertically planar inertial wave, following \citet{rocha2018stimulated}, we relate \textit{(i)} mean vertical vorticity to mean pressure and wave energy, and \textit{(ii)} wave-induced change of mean vorticity to wave energy using PV conservation.
We assess the accuracy of these  predictions in numerical solutions and find remarkably good agreement in spite of the complexity of the full three-dimensional Boussinesq model and of the fact that the strong scaling assumptions required by the theory are only marginally satisfied for the parameters chosen.  Our results illustrate  the value of wave-averaged theories for the analysis of geophysical flows in the context of three-dimensional nonlinear dynamics.

\begin{table}
 \begin{center}
  \begin{tabular}{ccccccccc}
    $f$ & $N$ & $a$ & $m$ & $\Ro$ & $\Bu$ & $E_0$ & $\nu_\mathrm{h} $  & $\nu_\mathrm{z}$    \\[5pt]
    
     200 & 1600 & 0.45 & 288 & 0.05 & 0.0038 & 1/2 & $5\times10^{-18}$ & $5\times10^{-23}$ \\
 \end{tabular}
 
  \caption{Simulation parameters. The horizontal domain size, $L = 2\pi$, determines the unit of length; the unit of time is defined so that the initial energy density of the NIW is $E_0=1/2$. { The vortex strength is such that the initial Eulerian domain-averaged mean-flow kinetic energy density  is $0.1$; the Gaussian vortex in \eqref{vort_profile} has maximum azimuthal velocity $0.32\Ro f a = 1.45$ at  $r = 1.13 a$.}}{\label{tab:sim_parameters}} 
 \end{center}
\end{table}

\section{Numerical solution of the Boussinesq equations}\label{sec:model}

We analyse solutions of the non-hydrostatic Boussinesq equations for an initial condition consisting of a NIW superimposed on a barotropic anticyclonic vortex with the initial vertical vorticity
\beq\label{vort_profile}
\zeta_0(r) = - \Ro f \, \e^{-r^2/a^2}\com 
\eeq
where $r$ is the radial coordinate, $a$ the vortex radius, and $\Ro$ the maximum Rossby number,
\begin{equation}
	\Ro = |\zeta_0|_{\mathrm{max}} / f. 
\end{equation}
The NIW is initially horizontally uniform, with vertical structure $\e^{\i m z}$ with vertical wavenumber $m$ and initial energy $E_0$. In addition to the Rossby number, the flow is characterised by the Burger number
\begin{equation}
	\Bu   = {N}^2/({f m a} )^2,
\end{equation}
where the inverse wavenumber $m^{-1}$ and  vortex radius $a$ are respectively used to define the characteristic vertical and horizontal scales.  We focus on the regime where both $\Ro$ and $\Bu$ are small as required for nearly geostrophic  mean dynamics and NIW dynamics \citep{young1997propagation}.

The Boussinesq equations are solved in a triply periodic domain, using a code adapted from that of \cite{waite2006transition}, which relies on a de-aliased pseudospectral method and a third-order Adams--Bashforth scheme with time step $0.06/f$. The $(2 \pi)^2 \times 2 \pi /36$ domain is discretised  using a  $1152^2 \times 96$ uniform grid. A hyperdissipation of the form $\nu_\mathrm{h} (\partial_x^2 + \partial_y^2)^4 +\nu_\mathrm{z} \partial^8_z$  is used for the momentum and density equations.  
The simulation parameters  are listed in  table \ref{tab:sim_parameters}.
The domain size is much larger than the vortex radius (${2\pi}/a \approx 14$) to mitigate the effect of waves re-entering from the boundaries. The NIWs are initialised with a horizontal velocity 
\beq \label{eq:wavIC}
(u'_0,v'_0) = (\cos m z , \sin mz)\, ;
\eeq
thus the initial  NIW kinetic energy density is $E_0 = (u_0^2+v_0^2)/2=1/2$. The vortex strength is such that the Eulerian mean-flow kinetic energy density  is $0.1$.

\begin{figure}
    \centering
     \includegraphics[width=1\linewidth]{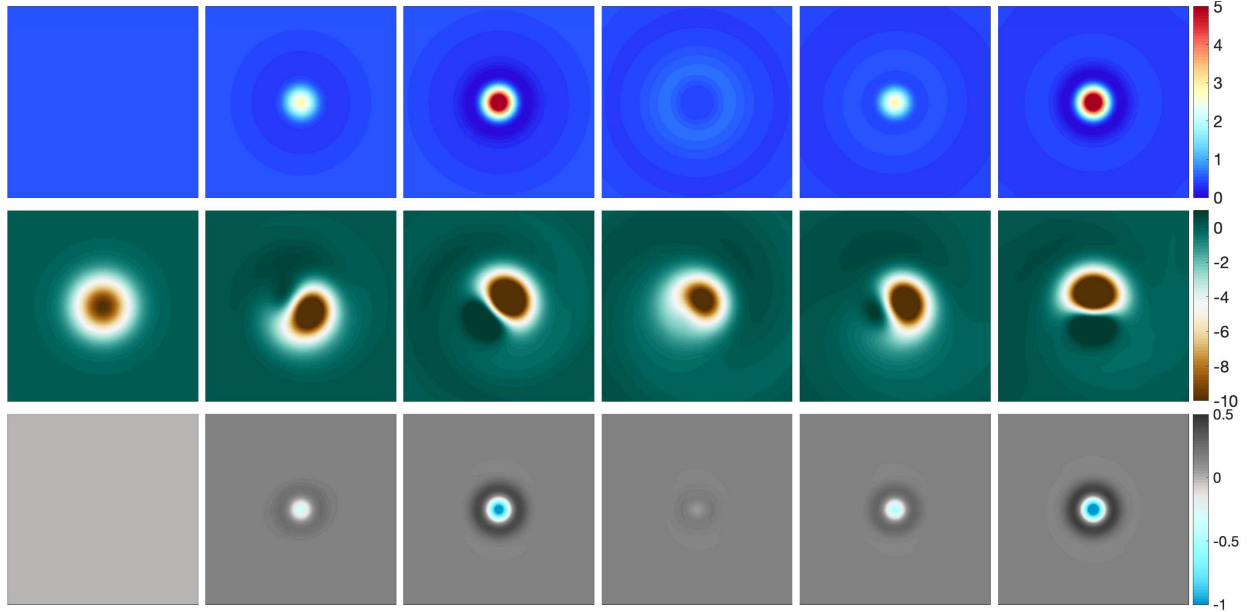}
\caption{ Horizontal slices of wave kinetic energy $({u'}^2 + {v'}^2)/2$ (top row), vertical vorticity $\zeta$ (middle row) and change in Eulerian-mean vertical vorticity $\overline{\zeta}-\zeta_0$ (bottom row) at times (from left to right) $t=0$, 20.4, 43.2, 76.3, 99.2 and 122.1 inertial periods $2\pi/f$.}
\label{fig:horizontalslices}
\end{figure}

The evolution of the NIW in the presence of a vortex is illustrated  by the sequence of snapshots of horizontal slices shown in figure \ref{fig:horizontalslices}. The NIW kinetic energy $({u'}^2+{v'}^2)/2$ is displayed in the top row, the vertical vorticity in the middle row, and the (barotropic) mean vorticity in the bottom row (this is also the Eulerian mean vorticity). The wave-energy concentration is modulated periodically in time, with a period much larger than the inertial period. The dynamics of the vertical vorticity $\zeta$ is dominated by the fast oscillation at the inertial frequency induced by NIW advection of $\overline{\zeta}$. This oscillation is filtered out by vertical averaging, leaving an  Eulerian mean vorticity $\overline{\zeta}$ that is almost axisymmetric and oscillates slowly in unison with the wave energy.

In the top row of  figure \ref{fig:horizontalslices} the NIW  energy level in the vortex core increases by over a factor of five above that of the initial condition \eqref{eq:wavIC}. This  focussing by the mean flow  results  in  an approximately axisymmetric trapped structure; this is a finite-amplitude, near-inertial normal mode of the vortex \citep{llewellyn1999near}. In anticipation of the strong wave expansion in \eqref{eq:expansion} below, we emphasize that this concentration of NIW energy happens even if   mean-flow velocity is much less  than NIW velocity:  weak mean flows shift the  lowest frequency of the internal wave band to $f + \overline{\zeta}/2$  \citep{Kunze1985,young1997propagation}. Thus vortices with $\overline{\zeta}/2<0$ support trapped  modes with frequencies slightly less than $f$, and the initial condition \eqref{eq:wavIC} projects onto this mode. We now turn to the mean-flow effects of this trapped mode.

\section{Wave-averaged analysis}\label{sec:balance}

We first derive the wave-averaged geostrophic balance \eqref{eq:averagedgeostrophy} and wave-averaged PV conservation for general inertia-gravity waves, then consider their application to NIWs and the numerical solution of \S\ref{sec:model}.

\subsection{{The strong-wave expansion}} \label{sec:waveaveragedgeostrophy}

 We provide only an informal sketch of the derivation and refer the reader to \citet{wagner2015available} for details (and  \citet{thomas2018wave} for related shallow-water results). We start with the Boussinesq equations
\begin{subequations} \label{eq:boussinesq}
\begin{align}
 \bm{u}_t + \bm{u}\bcdot \grad \bm{u}  + f \bm{z} \times \bm{u} + \grad p&=  b \bm{z}\com \label{eq:momentum} \\ 
b_t + \bm{u} \bcdot \grad b + N^2 w &=0\com  \\
\grad \bcdot \bm{u} &=0\per
\end{align}
\end{subequations}  For simplicity we take the Brunt--V\"ais\"al\"a frequency $N$ constant (see \citet{wagner2015available} for the non-constant case). 

We  assume that the flow consists of fast waves, with small amplitude $\eps \ll 1$, interacting with a slow flow, with smaller amplitude $\eps^2$. This is the \textit{strong-wave} assumption which makes it possible to capture the impact of the waves on the flow without the need to carry out the perturbation expansion to an unwieldy high order. The flow and wave amplitudes are assumed to vary on the  slow timescale of quasigeostrophic dynamics, which is $O(\eps^{-2} f)$ in our notation.
We expand all dynamical fields as
\begin{equation} \label{eq:expansion}
\bm{u} = \eps \bm{u}' + \eps^2 \overline{\bm{u}} + \cdots
\end{equation}
Here the prime identifies the wave component, the overbar the Eulerian-mean component obtained by averaging over the fast wave timescale, and the dots include a rapidly varying $O(\eps^2)$ term as well as $O(\eps^3)$ terms. Introducing \eqref{eq:expansion} into \eqref{eq:boussinesq} gives, at order $\eps$, the linear inertia-gravity-wave equations
\begin{subequations} \label{eq:IGWs}
\begin{align}
\bm{u}'_t + f \bm{z} \times \bm{u}' + \grad p' &=  b' \bm{z}\com
 \label{eq:IGWsmom} \\
b'_t + N^2 w' &=0 \com \label{Stoksey} \\
\grad \bcdot \bm{u}' &=0\per \label{eq:bw}
\end{align}
\end{subequations}

Averaging the next-order equations gives
\begin{subequations} \label{eq:meanflow}
\begin{align}
\overline{\bm{u}' \bcdot \grad \bm{u}'} + f \bm{z} \times\overline{ \bm{u} }+ \grad \overline{p} &= \overline{b} \bm{z} \com \label{eq:meanflowmom} \\
\overline{\bm{u}' \bcdot \grad b'} + N^2 \overline{w} &=0\com \label{eq:meanbuoyancy} \\
\grad \bcdot \overline{\bm{u}} &=0\, . \label{divmean}
\end{align}
\end{subequations} 
We can rewrite the nonlinear terms in \eqref{eq:meanflowmom} in terms of the wave displacement $\bm{\xi}'$. For instance, we have
\begin{subequations} \label{passage}
\begin{align} \label{eq:r1}
\overline{\bm{u}' \bcdot \grad \bm{u}'} &= - \overline{\bm{\xi}' \bcdot \grad \bm{u}'_t} \, , \\
&= f\bm{z} \cross \overline{\bm{\xi}' \bcdot \grad \bm{u}'} + \overline{(\bm{\xi}' \bcdot \grad) \grad  p'} - \overline{(\bm{\xi}' \bcdot \grad)b} \,  \bm{z} \, , \label{pass7}\\
&= f \bm{z} \cross  \barS{\bm{u}} +   \grad \half  \barS{p} - \barS{b} \bm{z}\, . \label{pass11}
\end{align}
\end{subequations} 
In passing from \eqref{pass7} to \eqref{pass11} we have used  the remarkable result
\beq \label{eq:remarkable}
(\overline{\bm{\xi}' \bcdot \grad) \grad p'} = \tfrac{1}{2} \grad (\overline{\bm{\xi}' \bcdot \grad p'}) = \tfrac{1}{2} \grad \barS{p},
\eeq
where $\barS{p} = \overline{\bm{\xi}' \bcdot \grad p'}$ is the Stokes pressure. The identity \eqref{eq:remarkable}  is established in \citet{wagner2015available} and a simplified proof is given in Appendix \ref{app:A}. 
Using  \eqref{pass11}  the mean momentum equations in \eqref{eq:meanflowmom} are reduced  to the wave-averaged balance equations \eqref{eq:averagedgeostrophy} with
\begin{equation} \label{eq:barLb}
\barL{b} = \overline{b} + \barS{b} \quad \textrm{and} \quad 
\overline{\pi} = \overline{p} + \tfrac{1}{2} \barS{p}, 
\end{equation}
where $\barS{b}= \overline{\bm{\xi}' \bcdot \grad b'}$ is the Stokes buoyancy. 
Finally,  with $\overline{\bm{u}'\bcdot \grad b'}= - \overline{\bm{\xi}'\bcdot \grad b'_t} = N^2 \overline{\bm{\xi'} \bcdot \grad w'}$,  one finds from \eqref{eq:meanbuoyancy}  that $\barL{w}  = \overline{w} + \barS{w} = 0$.



It is natural to introduce the streamfunction $\barL{\psi} = \overline{\pi}/f$ to write the components of the wave-averaged balance relations \eqref{eq:averagedgeostrophy} as
\begin{equation}\label{eq:averagegeostrophy}
\barL{\bm{u}} = (-\barL{\psi}_y, \barL{\psi}_x,0) \quad \textrm{and} \quad \barL{b} = \barL{\psi}_z,
\end{equation}
mirroring the familiar geostrophic and hydrostatic relations {of  quasigeostrophy}. The vorticity of the Lagrangian-mean flow is then
\begin{equation} \label{eq:barLzdeta}
\barL{\zeta} = \barL{v}_x - \barL{u}_y = \Dlt \barL{\psi},
\end{equation}
where $\Dlt = \partial_x^2 + \partial_y^2$ is the horizontal Laplacian. We emphasise that $\barL{\psi}$ and $\barL{\zeta}$ are the streamfunction and vorticity corresponding the Lagrangian-mean velocity $(\barL{u},\barL{v})$ but are not themselves Lagrangian means of any specific fields. This is evident from the  factor $1/2$ in \eqref{eq:barLb} and the fact that $\barL{\zeta} \not= \overline{\zeta} + \overline{\bm{\xi}' \bcdot \grad \zeta'}$ (the difference is related to the curl of the pseudomomentum, see e.g.\ \citet{buhler2014waves} or \citet{gilbert2018geometric}).

\subsection{Wave-averaged balance with  near-inertial waves}


The horizontal velocity $(u',v')$ of the vertically planar NIWs can be
written in the complex form
\begin{equation} \label{eq:NIW}
	u'(x,y,z,t) + \i \, v'(x,y,z,t) = \phi(x,y,t) \, \e^{\i (m z  - f t)},
\end{equation} 
where $\phi$ is a slowly-varying complex amplitude,  known as the back-rotated velocity. The form  in \eqref{eq:NIW} is an excellent approximation for the waves in our simulation. It makes clear that vertical averaging is equivalent to time averaging over the fast wave timescale. The wave kinetic energy is
$({u'}^2 + {v'}^2)/2=|\phi|^2/2$ and thus the top row of figure \ref{fig:horizontalslices} shows the evolution of  $|\phi|^2/2$. The back-rotated velocity $\phi$ evolves on the slow time scale $ \eps^{-2} t$, where $\eps$ is the order parameter in the strong-wave expansion \eqref{eq:expansion}. This long-term evolution   is obtained  by proceeding beyond the leading-order wave equation in \eqref{eq:IGWs} so that mean-flow effects, such as advection and $\zeta$-refraction, are revealed in the YBJ equation \citep{AY2019,AY2020,young1997propagation}.


\begin{figure} 
    \centering
    \begin{minipage}{.5\linewidth}
         \centerline{\includegraphics[trim=1cm 0 .5cm 0, width=1\linewidth]{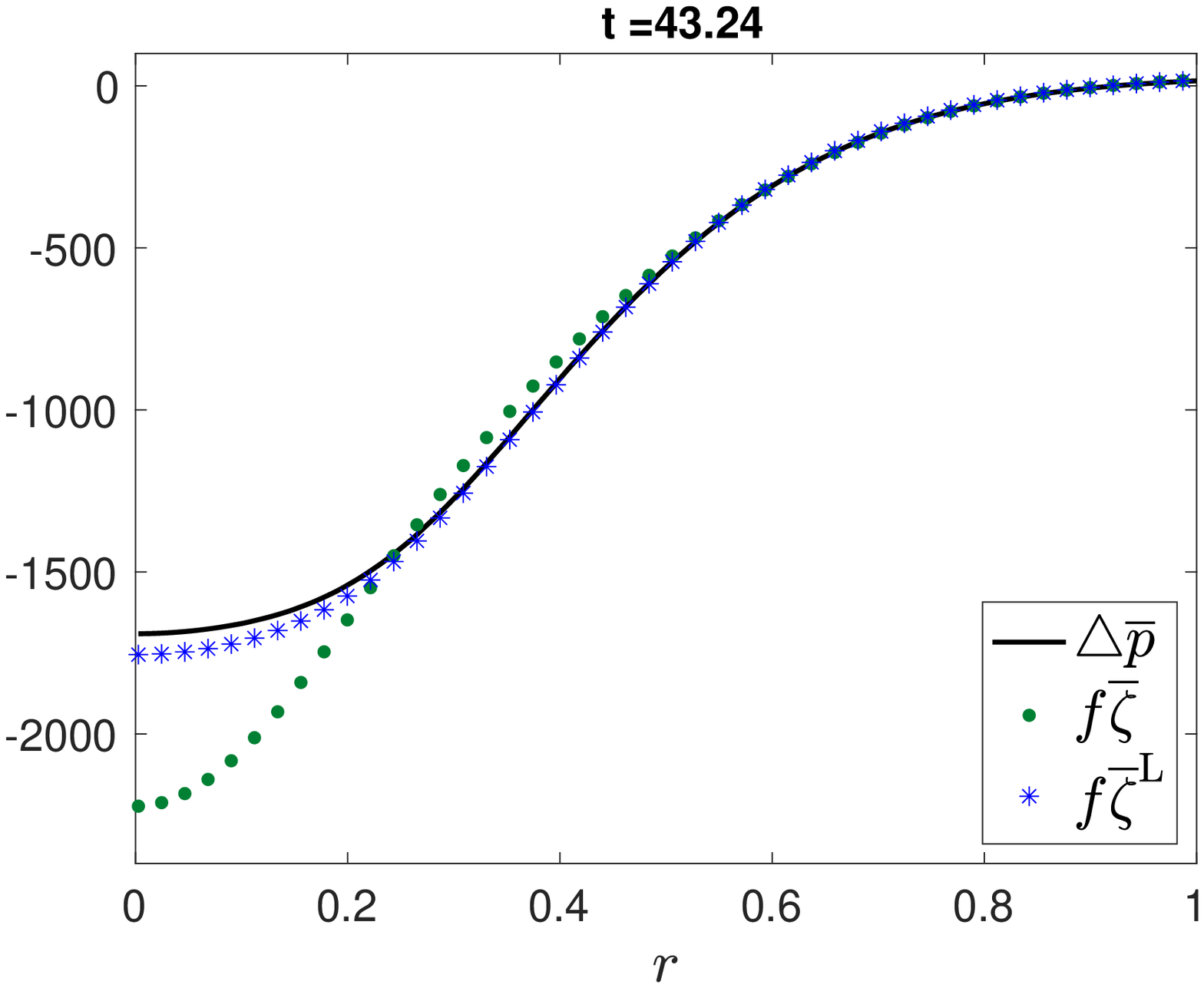}}
    \end{minipage}%
    \begin{minipage}{.5\linewidth}
        \centerline{\includegraphics[trim=1cm 0 .5cm 0, width=1\linewidth]{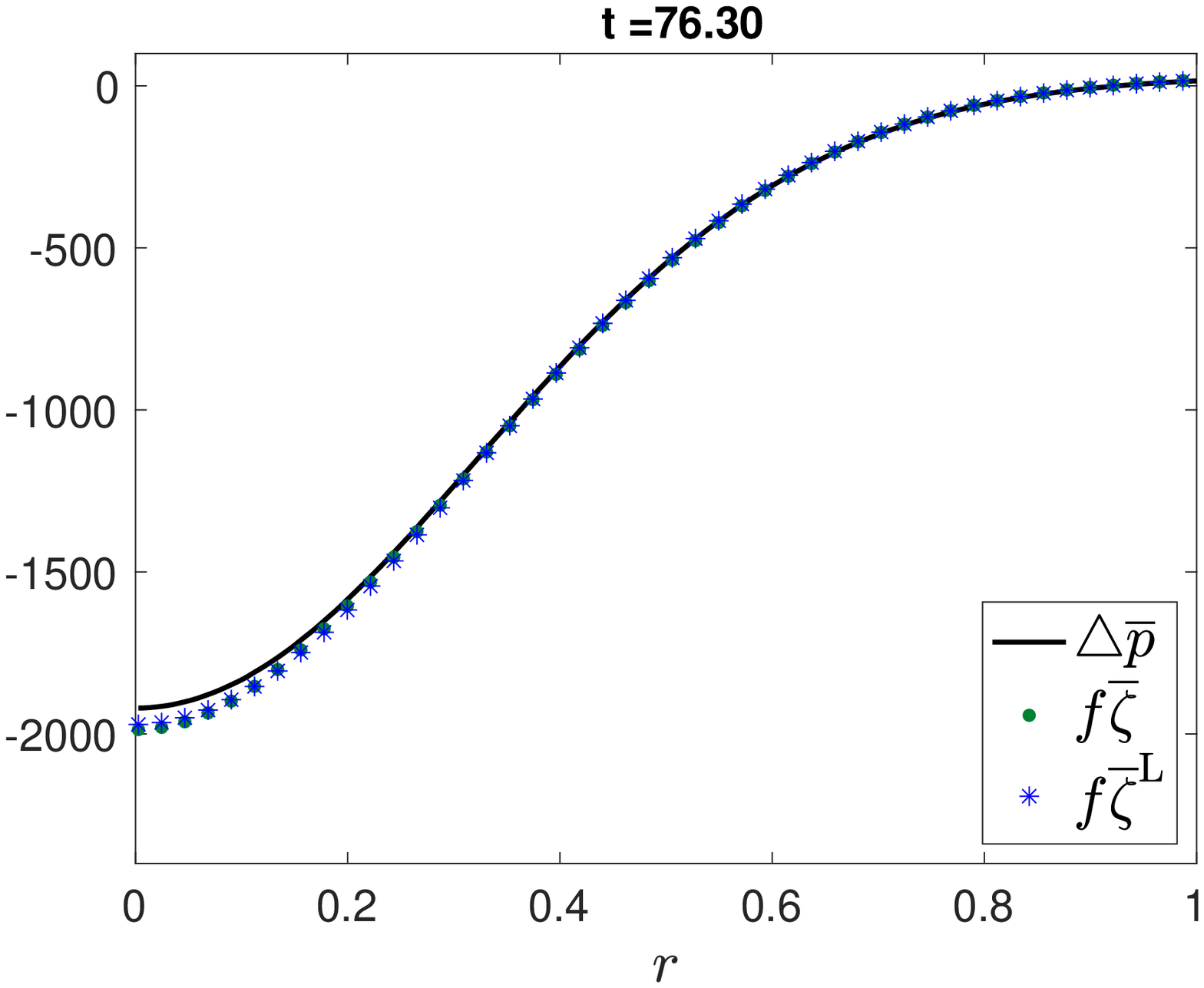}}
    \end{minipage}%
 \caption{Radial profiles of $f$ times the Lagrangian- and Eulerian-mean vertical vorticity, and of  $\Dlt \overline{p}$.  The times $t = 43.2$ (left) and $76.3$ (right) inertial periods correspond to maximally concentrated and nearly uniform wave energy respectively.} \label{fig:geo}    
     
\end{figure}

Using the back-rotated velocity $\phi$ dramatically simplifies the wave-averaged balance  \eqref{eq:averagedgeostrophy}. First, $p' \approx 0$ for NIWs, so $\overline{\pi} = \overline{p}$ and  taking the horizontal divergence of \eqref{eq:averagedgeostrophy} gives
\begin{equation} \label{eq:zetabarL}
f \barL{\zeta} = \Dlt \overline{p}.
\end{equation}
Second, for vertically planar NIWs the horizontal part of the Stokes velocity can be computed  as
\begin{equation} \label{eq:ubarLA}
(\barS{u},\barS{v})=(\mathcal{A}_y,-\mathcal{A}_x), \quad \textrm{where} \ \ \mathcal{A} = |\phi|^2/(2f)
\end{equation} 
is the wave action density, whose negative plays the role of a streamfunction for the Stokes velocity \citep[see][and Appendix \ref{app:A}]{rocha2018stimulated}. Combining  \eqref{eq:zetabarL} and \eqref{eq:ubarLA} relates the Eulerian-mean vertical vorticity to the pressure according to
\begin{equation} \label{eq:diagnostic}
f\, \left( \overline{\zeta} - \Dlt \mathcal{A} \right)= \Dlt \overline{p}.
\end{equation}
This is the form of wave-averaged geostrophy that we test in our numerical simulation. Figure \ref{fig:geo} shows radial profiles (obtained by azimuthal averaging) of the Eulerian-mean vorticity $\overline{\zeta}$, vorticity of the Lagrangian-mean flow $\barL{\zeta}$ and horizontal Laplacian of mean pressure $\Dlt \overline{p}$ evaluated from the simulation. Figure \ref{fig:geo}  shows snapshots at   two times corresponding to a maximum in the wave-energy concentration (left) and to a nearly uniform wave field (right).  It is clear from  that when the wave energy is non-uniform (left), geostrophic balance holds to a much better accuracy in the wave-averaged sense \eqref{eq:zetabarL} than in the conventional sense: it is  the Lagrangian-mean velocity, rather than the Eulerian-mean velocity, that is in geostrophic balance.

We have verified that the small  difference between $\barL{\zeta}$ and $\Dlt \overline{ p}$ in figure~\ref{fig:geo} results from the cyclostrophic correction to vortex geostrophy (not shown):  one restriction on the utility  of wave-averaged balance is that the  Stokes--Coriolis force, proportional to $f \barS{\bm{u}}$,  should be larger than non-wave ageostrophic terms such as the
aforementioned cyclostrophic correction. The formal justification of this restriction is in \eqref{eq:expansion}: the Stokes--Coriolis force is of order $\eps^2$ while the non-wave ageostrophic terms are of order $\eps^4$. For  the numerical  solution summarized in table \ref{tab:sim_parameters} there is not a  large difference between the mean and  NIW kinetic energies e.g. the maximum vortex velocity is only a factor of about  two  smaller than the maximum  NIW velocity  attained at $t = 43.2$. Nonetheless, in this numerical solution  the wave-averaged correction to geostrophic balance is much more important than cyclostrophic effects and other higher order non-wave corrections to balance.

\subsection{{Potential vorticity}}

We can go beyond the diagnostic relation \eqref{eq:averagegeostrophy} and make dynamical predictions using the conservation of {potential vorticity (PV hereafter)}. After removing the constant contribution $N^2 f$ and scaling by $N^2$, the PV can be written as
\begin{equation}
q = v_x - u_y + f b_z/N^2 + \left( \grad \times \bm{u} \right) \bcdot \grad b/N^2.
\end{equation}
Because the linearised inertia-gravity wave equations \eqref{eq:IGWs} imply that $v'_x - u'_y + f b'_z/N^2=0$, the PV is an $O(\eps^2)$ quantity. Its material conservation $\partial_t q + \bm{u} \bcdot \grad q =0$ shows that its leading-order approximation 
is evolving on the slow, O($\eps^{-2} f^{-1})$ time scale  and hence that it can be approximated by its time average. Using this we obtain, to leading order,
\begin{equation} \label{eq:PV1}
q = \barL{\zeta} + {f^2} \barL{\psi}_{zz} /{N^2} \underbrace{ - \barS{v}_x + \barS{u}_y - f \barS{b}_z/N^2 +  \overline{\left( \grad \times \bm{u}' \right) \bcdot \grad b'}/N^2}_{q^{\mathrm{w}}},
\end{equation}
where  \eqref{eq:averagegeostrophy} has been used. 
The first two terms on the right of \eqref{eq:PV1}  make up the familiar quasi-geostrophic PV, here involving the streamfunction of the Lagrangian-mean flow. The remaining terms constitute the wave PV denoted $q^{\mathrm{w}}$. We note that the wave-averaged geostrophic balance and PV conservation under advection by $\barL{\bm{u}}$ also follow directly from GLM theory \citep{andrews1978exact,buhler1998non-dissipative,holmes2011particle,xie2015generalised,gilbert2018geometric}.

\begin{figure}
\includegraphics[width=1\linewidth]{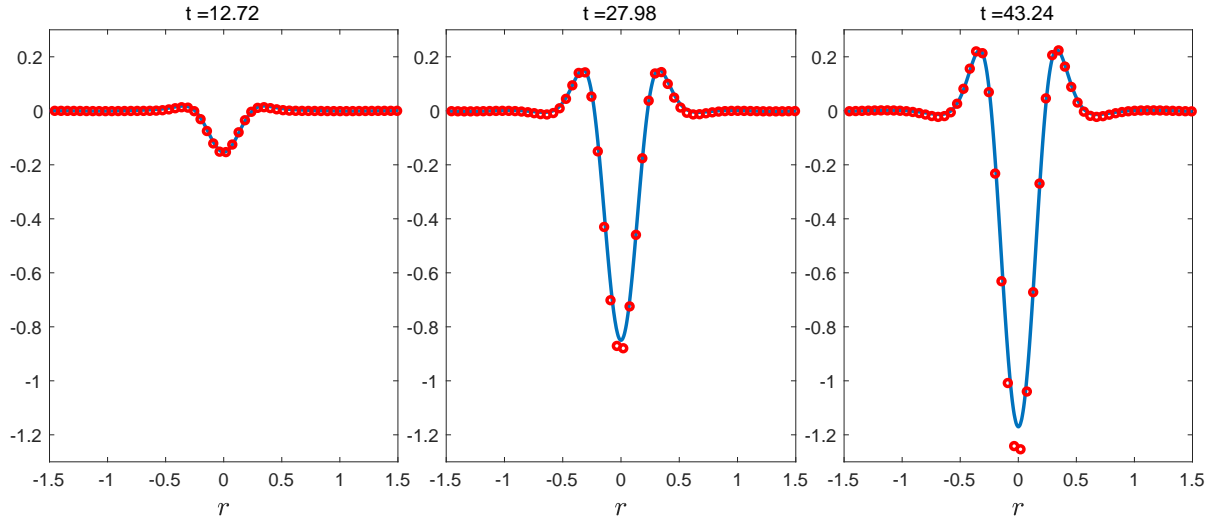}
\caption{Comparison of mean-vorticity change $\overline{\zeta}-\zeta_0$ (\textcolor{red}{$\circ$}) with $\Dlt \mathcal{A}/2$ (solid line) along a radial line and for $t=12.7$ (left), 30 (middle) and 43.2 (right) inertial periods.}
\label{fig:vorticityaction}
\end{figure}

For NIWs, $q^{\mathrm{w}}$  can be expressed in terms of the back-rotated velocity $\phi$;  
retaining only leading-order terms in the small Burger number the result is
\begin{equation} \label{eq:Q0}
q = \barL{\zeta} + {f^2} \barL{\psi}_{zz} /{N^2} + \Dlt |\phi|^2/(4f) + \i \partial(\phi,\phi^*)/(2f),
\end{equation}
where $\partial(\cdot,\cdot)$ denotes the Jacobian operator \citep[see][]{xie2015generalised,wagner2016three}. For an axisymmetric wave field $\phi(r,t)$ and barotropic mean flow \eqref{eq:Q0}  reduces  to
\beq
q = \barL{\zeta} + \Dlt \mathcal{A}/2\com
\eeq
where $\Dlt(\cdot) = r^{-1} \partial_r(r \partial_r \cdot)$. For waves that are initially uniform in the horizontal, \WRY{as in the initial condition \eqref{eq:wavIC}}, the Stokes drift and horizontal Laplacian vanish initially, so the conservation of $q$ -- which holds pointwise since $q$ is axisymmetric and $\barL{\bm{u}}$ has no radial component -- implies that
\beq \label{eq:zetaL}
\barL{\zeta}  = \zeta_0 -  \Dlt \mathcal{A}/2\com 
\eeq 
where $\zeta_0(r)$ is the initial Gaussian vertical vorticity in \eqref{vort_profile}. Using the expression for the Stokes velocity in  \eqref{eq:ubarLA}, the Eulerian-mean vorticity is 
\beq \label{eq:zeta}
\overline{\zeta} = \zeta_0 + \Dlt \mathcal{A}/2.
\eeq
This result explains the correlation between the evolution of the wave kinetic energy $({u'}^2 + {v'}^2)/2 = f \mathcal{A}$ and $\overline{\zeta}$ observed in figure \ref{fig:horizontalslices}. We test the prediction in \eqref{eq:zeta} quantitatively in figure \ref{fig:vorticityaction} by comparing the change in mean vorticity, $\overline{\zeta}-\zeta_0$, with $\Dlt \mathcal{A}/2$ at three different times corresponding to different phases in the oscillation of wave energy. The good match between the two quantities confirms the validity of \eqref{eq:zeta}. Thus  conservation of wave-averaged PV predicts the dynamics of the mean flow in the presence of waves with substantial amplitude.

\section{Discussion}

This paper examines the way geostrophic balance -- a cornerstone of geophysical fluid dynamics -- is altered in the presence of inertia-gravity waves to become a balance between the Coriolis force associated with the Lagrangian-mean velocity and a wave-modified mean pressure. The wave-induced correction to geostrophy has long been  known  \citep[e.g.][]{moore1970mass} but it has received much less attention than the  issue of \WRY{non-wave}  finite-Rossby-number corrections associated with higher-order balance  \citep[e.g.][]{machenhauer1977dynamics,leith1980nonlinear,allen1993intermediate,warn1995rossby,mcintyre2000potential,mohebalhojeh2001hierarchies,vanneste2013balance,kafiabad2018spontaneous,kafiabad2017rotating}. By focusing on numerical simulations in a very simple setup, in which vertically planar NIW waves are superimposed on a barotropic vortex, we are able to show unambiguously that the wave-induced correction matters and, for NIWs, can be estimated from the wave energy, consistent with theoretical predictions,  at least in the simple case considered here. We further show that the material conservation of a wave-averaged PV can be exploited to predict mean-flow changes from the wave energy.  We emphasise that the concepts of wave-averaged balance and wave-averaged  PV are not limited to the present case of vertically planar NIWs propagating through a barotropic mean flow.  They apply in the presence of vertically sheared mean flows,  as  in  the numerical NIW simulations in \cite{AY2020},  \cite{ThomasArun2020} and \cite{ThomasDaniel2020}, and to generic inertia-gravity waves.
In  the case of baroclinic mean flows, however, the waves cannot be extracted by removing the vertical average as done here. The standard normal-mode decomposition \citep[e.g.][]{bartello1995geostrophic} also falls short for this task: in the case of strong waves, because of the importance of the Stokes--Coriolis force $f \bm{z} \times \barS{\bm{u}}$,  balanced flow is poorly approximated by the linearised part of PV, and an appropriate form of time filtering is required to separate waves from balanced flow.


To  assess the importance of wave-induced corrections to geostrophic balance in an oceanic context, we need to compare the size of the Stokes velocity induced by inertia-gravity waves to typical mean-flow velocities. We focus here on the case of NIWs in the ocean's mixed layer represented, as is standard, by a slab model \citep[e.g.][]{alford2016near}. We note that the relation $\barS{\bm{u}} = (-\mathcal{A}_y,\mathcal{A}_x)$ applies to (vertically independent) waves in a slab model as well as to the vertically planar waves considered so far. We estimate $\mathcal{A}$ from the values of inertial-wave kinetic energy inferred by 
\citet{chaigneau2008global} from drifter data. In parts of the ocean with strong wind forcing, they found this kinetic energy to be of the order of $10^3$ J m$^{-2}$. This corresponds to
$({u'}^2+{v'}^2)/2 = 10^3/(\rho h) \approx 10^{-2}$ m$^2$ s$^{-2}$, taking the water density $\rho=10^3$ kg m$^{-3}$ and mixed-layer depth $h=100$ m, and hence $\mathcal{A} \approx 10^2$ m$^2$ s$^{-1}$ for $f = 10^{-4}$ s$^{-1}$. Using a spatial scale of 10 km based on recent simulations of NIWs \citep{AY2020}, we finally estimate $|\barS{\bm{u}}| \approx 10^{-2}$ m s$^{-1}$, a substantial fraction of typical geostrophic velocities at the ocean surface. Larger values are obtained for smaller $f$ (lower latitudes) and a shallower mixed layer. 
The NIW Stokes velocity is also comparable in magnitude to the Stokes drift of  surface gravity waves. However, unlike the latter, {near-inertial Stokes drift} extends across the mixed layer and into the ocean interior, hence it is potentially more important for transport. The contribution from other types of inertia-gravity waves, including internal tides, also deserves attention.


Our order-of-magnitude estimate raises the possibility that wave-averaged corrections to geostrophic balance affect the inference of sea-surface geostrophic velocity from satellite altimetric measurements. Assuming that sea-surface-height measurements provide an approximation to the Eulerian-mean height and hence, for NIWs, also of the Lagrangian-mean height (since NIWs have little effect on the sea-surface height), the surface velocity inferred by standard geostrophy is a Lagrangian mean rather than an Eulerian mean as normally assumed. It would be useful to examine whether and how this difference can be accounted for.

We conclude by noting that, while the present paper concentrates on the relation between wave energy and wave-induced mean flow, it is also desirable to analyse the mechanism that leads to the slow wave-energy oscillations seen in figure \ref{fig:horizontalslices}. We leave this analysis for  another publication \citep{KVY2021}.

\medskip
\noindent
\textbf{Acknowledgments.} HAK and JV are supported by the UK Natural Environment Research Council grant NE/R006652/1. WRY is  supported by the National Science Foundation Award OCE-1657041. This work used the ARCHER UK National Supercomputing Service.

\medskip
\noindent
\textbf{Declaration of interests.} The authors report no conflict of interest.

\appendix

\section{Derivation details} \label{app:A}

We establish \eqref{eq:remarkable} by rewriting the wave momentum equation \eqref{eq:IGWsmom} as
\begin{equation}
\grad p' = \mathsf{L} \bm{\xi}', \quad \textrm{where} \ \ \mathsf{L}
= \begin{pmatrix}
- \partial_t^2 & f \partial_t & 0 \\
-f \partial_t & - \partial_t^2 & 0 \\
0 & 0 & -\partial_t^2 - N^2
\end{pmatrix}.
\end{equation}
The linear operator $\mathsf{L}$ is self-adjoint in the sense that $\overline{\bm{a} \bcdot \mathsf{L} \bm{b}} = \overline{\bm{b} \bcdot \mathsf{L} \bm{a}}$ for arbitrary time-periodic vectors $\bm{a}$ and $\bm{b}$. Using this we compute
\begin{equation}
\partial_i ( \overline{\bm{\xi}' \bcdot \grad p'}) =  \partial_i ( \overline{\bm{\xi}' \bcdot \mathsf{L} \bm{\xi}'}) =  2  \overline{\bm{\xi}' \bcdot \mathsf{L} \partial_i \bm{\xi}'} = 2  \overline{\bm{\xi}' \bcdot  \partial_i \grad p'} = 2  \overline{(\bm{\xi}' \bcdot  \grad) \partial_i p'}
\end{equation}
and \eqref{eq:remarkable} follows. 

We now turn to \eqref{eq:ubarLA}. The unaveraged Stokes velocity is
\begin{equation}
\ \ddxi  \bu '= \partial_t\big[ \half \ddxi \bxi' \big] - \half \curl \bh, 
\label{SD11}
\end{equation}
where $\bh =\bxi'\cross \bu'$ contributes to the absolute angular momentum density, and we used the standard vector identity for the curl of a cross product and that $\div \bm{\xi}' = \div \bm{u}' = 0$. The time average of \eqref{SD11} gives the explicit solenoidal representation of the Stokes velocity 
\begin{equation}
\barS{\bm{u}} = - \half \curl \overline{\bh}.
\label{eq:stokes17}
\end{equation}
For vertically planar waves, $\partial_z \overline{\bh} = 0$ hence
\begin{equation} \label{eq:barSuh}
(\barS{u},\barS{v})= \half (-\overline{h}_y,\overline{h}_x)
\end{equation}
where $\overline{h}= \overline{\xi'v' - \eta'u'}$  is the vertical component of $\overline{\bh}$.  For NIWs, $u'_t - f v' =v'_t + f u' = 0$, so $\overline{h}=(\overline{\xi' u'_t + \eta' v'_t})/f = - (\overline{{u'}^2 +{v'}^2})/f = - \mathcal{A}$ and \eqref{eq:barSuh} gives \eqref{eq:ubarLA}.

\bibliography{SLOB.bib}
\bibliographystyle{abbrvnat}

\end{document}